\documentclass[twocolumn,aps,prl,showpacs,amsmath,amssymb]{revtex4}
\usepackage{graphicx}% Include figure files

\begin{document}

\title{
Magnetoresistance of Untwinned YBa$_{2}$Cu$_{3}$O$_{y}$ 
Single Crystals in a Wide Range of Doping: Anomalous Hole-Doping 
Dependence of the Coherence Length}

\author{Yoichi Ando}
\author{Kouji Segawa}
\affiliation{Central Research Institute of Electric Power Industry, 
Komae, Tokyo 201-8511, Japan.}

\date{\today}

\begin{abstract}
Magnetoresistance (MR) in the $a$-axis resistivity of untwinned 
YBa$_{2}$Cu$_{3}$O$_{y}$ single crystals is measured for 
a wide range of doping ($y= 6.45 - 7.0$).  
The $y$-dependence of the in-plane coherence length $\xi_{ab}$ 
estimated from the fluctuation magnetoconductance indicates that 
the superconductivity is anomalously weakened in the 60-K phase; 
this gives evidence, together with the Hall coefficient and the 
$a$-axis thermopower data that suggest the hole doping to be 12\% 
for $y \simeq 6.65$, that the origin of the 60-K plateau is the 1/8 
anomaly.
At high temperatures, the normal-state MR data show signatures 
of the Zeeman effect on the pseudogap in underdoped samples. 
\end{abstract}

%74.25.Fy Transport properties
%74.40.+k Fluctuations 
%74.62.Dh Transition temperature variations; Effects of 
%          crystal defects, doping and substitution  
%74.72.Bk Y-based cuprates

\maketitle

Magnetoresistance (MR) is a useful tool to study the electron 
transport in metals, though its origin can be complicated.  
One can infer, for example, the significance of the 
spin-dependent mechanisms through the longitudinal MR, 
and the fluctuation magnetoconductivity (FMC) observed 
above $T_c$ of a superconductor can be used to estimate the 
coherence length. 
In high-$T_c$ cuprates, the normal-state orbital MR is 
known to violate the Kohler's rule and to 
show an unusual temperature dependence, $\sim (aT^2+b)^{-2}$ 
\cite{Harris}, 
which indicates that an unusual situation for the charge 
transport, possibly a scattering-time separation 
\cite{Anderson,Coleman}, is realized.  
Also, it has been discussed 
that a sizable FMC survives to much above $T_c$
in cuprates \cite{Harris,Kimura}, even twice as high as $T_c$ in 
optimally-doped La$_{2-x}$Sr$_x$CuO$_4$ (LSCO) \cite{Harris}.
Such unusual MR behavior naturally calls for detailed 
doping-dependence studies of the MR to better understand 
the charge transport in cuprates; however, 
there have been only a few reports on the 
hole-doping dependence of the MR behavior 
\cite{Harris,Kimura,Murayama,Amitin}, and the 
key issues such as the evolution of the role of spins 
or the evolution of the fluctuation contributions are 
not really understood yet.

In this Letter, we report the MR measurements of 
untwinned YBa$_{2}$Cu$_{3}$O$_{y}$ (YBCO) single 
crystals in a wide range of doping, from heavily underdoped 
($T_c$=20 K) to slightly overdoped regions.
Since YBCO contains Cu-O chains which can carry the electric 
current along the $b$-axis, we measure the magnetic-field 
dependence of the $a$-axis resistivity $\rho_a$ and pay 
particular attention to 
sorting out the genuine MR behavior of the CuO$_2$ planes 
with a careful analysis involving also the $b$-axis resistivity 
$\rho_b$ and the Hall resistivity $\rho_{\rm H}$. 
Based on the magnetoconductivity of the CuO$_2$ planes 
obtained after the analysis, we discuss both the normal-state 
magnetoconductivity at high temperatures and 
the FMC at lower temperatures.  
Most notably, the FMC is found to show a non-monotonic 
evolution with $y$ and its $y$-dependence suggests that the 
superconductivity is anomalously weakened in the ``60-K phase" \cite{60K} 
of YBCO; in combination with other 
in-plane transport properties which suggest that the hole doping 
is actually 1/8 at $y \simeq 6.65$, the present data give evidence 
for the 1/8-anomaly origin of the 60-K plateau.

\begin{figure}[b!]
\includegraphics[clip,width=8cm]{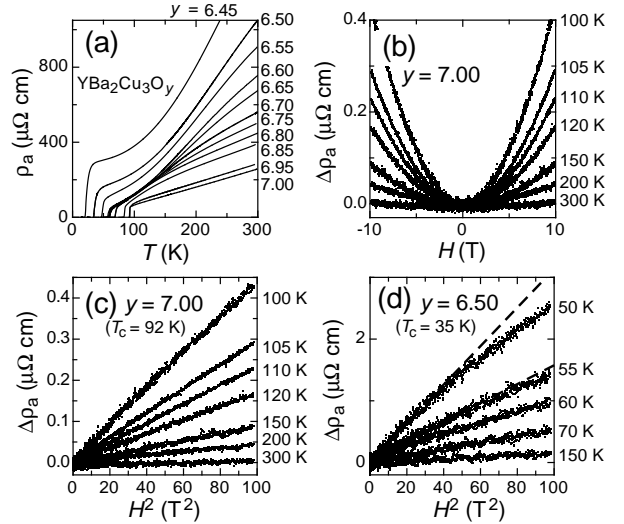}
\caption{\label{fig1}
(a) $\rho_a(T)$ data for all the $y$ values studied.
(b) Raw data of the transverse MR for $y$=7.0 at selected temperatures. 
(c) $\Delta \rho_a$ vs $H^2$ plots of the data in panel (b). 
(d) Transverse MR for $y = 6.50$ at selected temperatures; the dashed line 
is the linear fit to the low-field part of the data.}
\end{figure}

The YBCO single crystals are grown in Y$_2$O$_3$ crucibles by 
a conventional flux method \cite{Segawa}, and are carefully detwinned 
and annealed (see Ref. \cite{60K} for details); 
the crystals reported here are in the range of $y=6.45-7.0$
(the error in $y$ is $\pm$0.02).
The MR measurements are done using an ac four-probe 
method under sweeping magnetic fields of $\pm$10 T or $\pm$14 T. 
Both the transverse (magnetic field $H$ is perpendicular 
to the $ab$ planes) and the longitudinal ($H$ is parallel 
to the current $I$) MR are measured. 
The temperature is stabilized within $\sim$1 mK 
during the MR measurements by a home-build temperature 
regulation system employing both the Cernox resistance 
sensor and a capacitance sensor. 
Other details of the measurements of $\rho_a$, $\rho_b$, and 
$\rho_{\rm H}$ are described in Ref. \cite{60K}.

\begin{figure}[t!]
\includegraphics[clip,width=7cm]{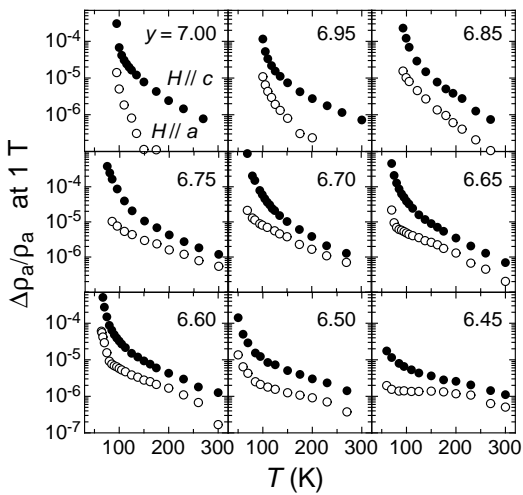}
\caption{\label{fig2}
$T$-dependences of the transverse ($\bullet$) and 
longitudinal ($\circ$) MR  
for most of the hole concentrations studied.}
\end{figure}

Figure 1(a) shows the $\rho_a(T)$ data for 
the series of our untwinned crystals. 
An unusual overlapping of the $\rho_a(T)$ data for the 
``60-K phase" samples, which is clear in Fig. 1(a), has 
been discussed in Ref. \cite{60K}.
Figure 1(b) shows examples of the raw MR ($\Delta \rho_a$) 
data for $y$=7.0.  All the data plotted in Fig. 1(b) 
show the ordinary $H^2$ dependence, which becomes clear in the 
plot of $\Delta \rho_a$ vs $H^2$ [Fig. 1(c)].
In the underdoped samples, however, we observed that the MR 
starts to become concave downward (or tends to saturate) in 
the $\Delta \rho_a$ vs $H^2$ plot upon approaching $T_c$, 
as shown in Fig. 1(d) for $y$=6.50 ($T_c$=35 K).  
Such $H$-dependence is most likely caused by a 
decrease in the characteristic magnetic-field scale to suppress 
superconducting fluctuations, and some vortex fluctuations in 
the pseudogap state \cite{Orenstein,Ong_Nernst} are possibly 
involved in this anomaly. 
In any case, when this anomalous $H$-dependence is observed 
near $T_c$, the MR data cannot be discussed on the same ground 
as those at higher temperatures, and therefore we do not include 
the MR data for temperatures very close to $T_c$ in the 
discussions afterwards.  However, as long as the temperature is 
\textit{not} very close to $T_c$, the low-field part of $\Delta \rho_a$ 
can reasonably be fitted with $H^2$ [as shown for the 50-K data 
in Fig. 1(d)] and for such data we determine the ``magnitude" of MR 
from the low-field slope of the $\Delta \rho_a$ vs $H^2$ plot.

Figure 2 shows the summary of the transverse 
and longitudinal MR ($[\Delta \rho_a/\rho_a]_{\perp}$ and 
$[\Delta \rho_a/\rho_a]_{//}$, respectively) 
for the whole doping range. 
In Fig. 2, one may notice that the longitudinal MR 
rapidly diminishes with increasing temperature in highly-doped 
samples ($y$=6.95 and 7.0), while it remains noticeable up to 
270 K in the underdoped samples.  While the low-temperature 
growth of the longitudinal MR is likely to originate from 
the Zeeman term in the FMC, its behavior in the 
high-temperature region is expected to reflect the properties 
of the normal-state. 
Figure 3 depicts the $y$ dependence of 
$[\Delta \rho_a/\rho_a]_{//}$ at high temperatures, 
where it is clear that there is a crossover near $y$=6.8 
above which the longitudinal MR is diminished. 

\begin{figure}[t!]
\includegraphics[clip,width=6cm]{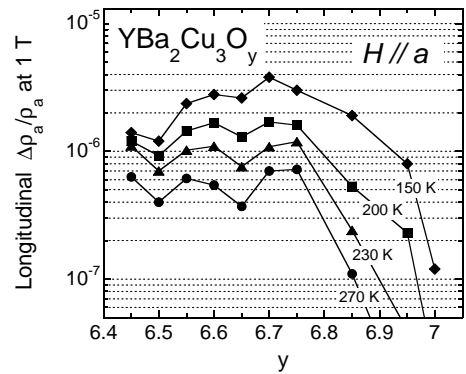}
\caption{\label{fig3}
$y$-dependence of the longitudinal MR at high temperatures 
where fluctuation contribution is negligible.}
\end{figure}

Since the longitudinal MR in cuprates mostly measures the Zeeman effect 
on the spin terms, the observed crossover in $[\Delta \rho_a/\rho_a]_{//}$ 
near $y$=6.8 is likely to be a manifestation of the Zeeman effect on the 
pseudogap which opens in the normal state of underdoped samples. 
Remember that the pseudogap is accompanied 
by a spin gap \cite{Timusk}, which reduces the 
magnetic scattering rate; since the magnetic scatterings are 
recovered when the spin gap is suppressed by the Zeeman effect, a 
measurable longitudinal MR is actually expected in the pseudogap state.

From the data of $[\Delta \rho_a/\rho_a]_{\perp}$ and 
$[\Delta \rho_a/\rho_a]_{//}$, we can deduce the orbital 
magnetoconductivity of the CuO$_2$ planes, although the correct 
procedure is not simple for an anisotropic system.
For an \textit{isotropic} system, the MR (up to terms in $B^2$) 
is written as 
$\Delta \rho/\rho = -\Delta \sigma/\sigma - (\sigma_{xy}/\sigma)^2$, 
where $\sigma$ is the conductivity and $\sigma_{xy}$ is the 
Hall conductivity \cite{Harris}. 
This formula can be generalized to an anisotropic system as 
$\Delta \rho_a/\rho_a = -\Delta \sigma_a/\sigma_a - 
\sigma_{xy}^2/(\sigma_a\sigma_b)$; 
note here that in an anisotropic system $\sigma_{xy}$ and 
$\rho_{\rm H}$ are related (up to terms in $B$) by 
$\sigma_{xy} = \rho_{\rm H}/(\rho_a\rho_b)$ \cite{60K}. 
Therefore, to obtain the correct magnetoconductivity from 
the MR data of an anisotropic system, one should use the formula 
$\Delta \sigma_a/\sigma_a = - \Delta \rho_a/\rho_a - 
\rho_{\rm H}^2/(\rho_a\rho_b)$.
This means one needs to measure not only the $a$-axis transport 
but also $\rho_b$ and $\rho_{\rm H}$. 
To the best of our knowledge, this correct formula has never been 
used for the analyses of the magnetoconductivity in cuprates, 
even though $\Delta \rho_a/\rho_a$ and 
$\rho_{\rm H}^2/(\rho_a\rho_b)$ can become the same order at 
high temperatures.
To obtain the orbital magnetoconductivity for our samples, 
we calculate both the transverse and the longitudinal 
magnetoconductivity by using the correct formula 
and take the difference.  The data of $\rho_b$ and $\rho_{\rm H}$ 
used in the analysis are shown elsewhere \cite{60K,Hall}.

\begin{figure}[t!]
\includegraphics[clip,width=7.5cm]{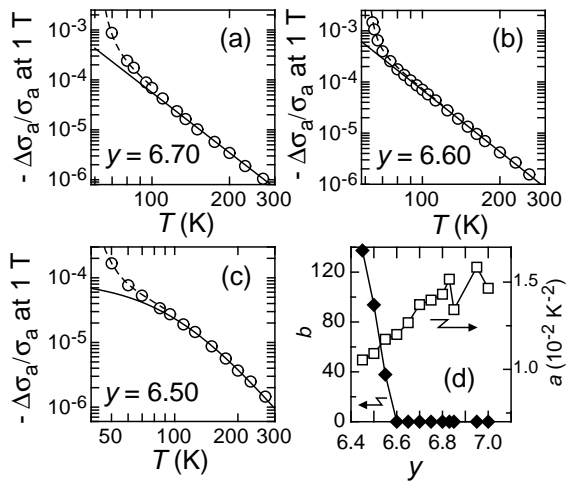}
\caption{\label{fig4}
$T$-dependences of the orbital magnetoconductivity 
for (a) $y$ = 6.70, (b) 6.60, and (c) 6.50.
The solid lines are the fits of the high-temperature region 
to the normal-state contribution $(aT^2+b)^{-2}$.
The parameters $a$ and $b$ are shown in panel (d).
The dashed lines in (a)-(c) are the fits of the data to the 
additional Aslamazov-Larkin orbital term.}
\end{figure}

Figures 4(a)-(c) show examples of the orbital magnetoconductivity 
data thus obtained.  
In the analysis of these data, 
we follow the recent trend \cite{Harris,Kimura,Semba} to interpret 
the high-temperature part to come from the normal-state contribution 
[which is expressed as $(aT^2+b)^{-2}$] rather than 
the Maki-Thompson orbital FMC \cite{MT}. 
The solid lines in Figs. 4(a)-(c) are the fits of the 
data to the normal-state contribution.  
The data at high temperatures are actually very well fitted with the 
$(aT^2+b)^{-2}$ dependence.  The offset $b$ is zero for $y$ of 
down to 6.60, but becomes non-zero for smaller $y$, which 
corresponds well to the behavior of the $\rho_a(T)$ data 
where a noticeable residual resistivity becomes apparent for 
$y < 6.60$ [Fig. 1(a)]; this seems to give additional support to the 
validity of this analysis.  
The $y$-dependence of the fitting parameters are summarized 
in Fig. 4(d). 
We note that the magneto-\textit{resistance} is not as well fitted 
with $(aT^2+b)^{-2}$ as the magneto-\textit{conductance}, which 
implies the necessity of the correct calculation for 
$\Delta \sigma_a/\sigma_a$. 

\begin{figure}[t!]
\includegraphics[clip,width=8cm]{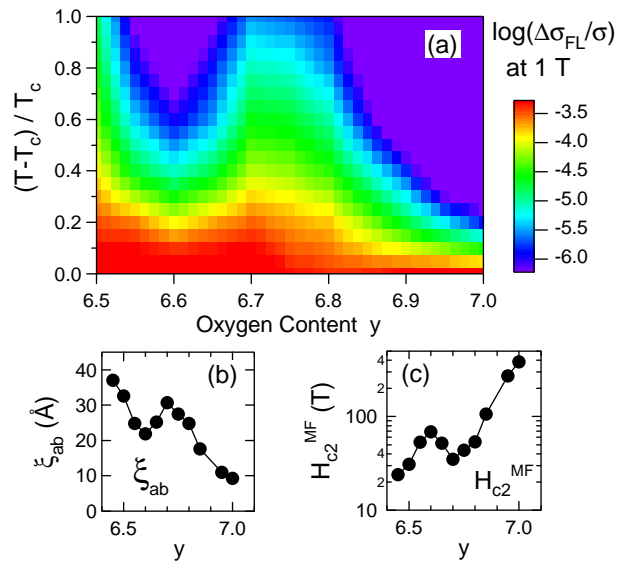}
\caption{\label{fig5}
(a) Evolution of the fluctuation magnetoconductivity 
in the $(T-T_c)/T_c$ vs $y$ plane. (b) $y$-dependence of 
$\xi_{ab}$ obtained from the ALO fits. (c) Mean-field 
$H_{c2}$ calculated from $\xi_{ab}$.}
\end{figure}

We can estimate the FMC by subtracting the normal-state 
contribution (determined from the high-temperature data) from the 
total magnetoconductivity.  
Figure 5(a) depicts the evolution of the FMC in the 
$(T-T_c)/T_c$ vs $y$ plane using a color scale. 
It becomes clear in Fig. 5(a) that, in addition to the general 
growth of the FMC with decreasing $y$, 
there is an enhancement in the FMC near $y$=6.7, which 
corresponds to the 60-K phase of YBCO; 
this means that the superconducting fluctuations are more easily 
suppressed with magnetic field and thus the characteristic 
field scale for superconductivity is reduced near $y$=6.7. 

This situation can be more quantitatively analyzed by fitting the 
FMC data to the Aslamazov-Larkin orbital (ALO) term \cite{Aronov}; 
the dashed lines in Figs. 4(a)-(c) are the fits of 
the lower-temperature data to the normal-state contribution 
plus the ALO term.  In determining the dashed lines, there are 
two fitting parameters, the in-plane and $c$-axis coherence lengths, 
$\xi_{ab}$ and $\xi_c$.  ($T_c$ is set to the zero-resistance $T_c$.)  
Since the fits become insensitive to 
$\xi_c$ for $y < 6.9$ where $\xi_c \alt 1$ \AA, 
for underdoped samples we fix $\xi_c$ to be near 1 \AA \ \ 
and only change $\xi_{ab}$ in the fits.
Although not many data points are available for fitting for 
each composition, we can determine $\xi_{ab}$ with roughly 
20\% error from the ALO fits. 

Figure 5(b) shows the $y$-dependence of $\xi_{ab}$ obtained 
from the ALO fits.  
The general trend is that $\xi_{ab}$ increases with decreasing 
hole doping, which is natural because the mean-field 
upper critical field at zero temperature, $H_{c2}^{\rm MF}$ 
[$=\Phi_0/(2\pi\xi_{ab}^2)$], 
is expected to be reduced as $T_c$ goes down.
However, for $y \simeq 6.7$ there is a marked anomaly that 
$\xi_{ab}$ deviates upwardly from the general trend, 
which corresponds to a \textit{suppression} of $H_{c2}^{\rm MF}$ 
in the 60-K phase [Fig. 5(c)].
This is in accord with the trend already apparent in Fig. 5(a).
We note that this implication on the upper critical field 
is actually corroborated by the behavior of the resistive transition 
in 16 T, where the 60-K-phase samples show a marked broadening compared 
to other compositions, as demonstrated in Fig. 6(a). 
Therefore, both $\xi_{ab}$ and the resistive transition 
appear to indicate that the superconductivity is anomalously 
weak to the applied magnetic field in the region near $y$=6.7.

\begin{figure}[t!]
\includegraphics[clip,width=8.5cm]{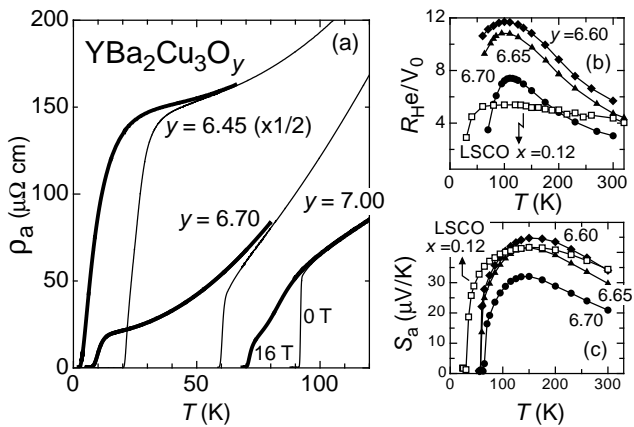}
\caption{\label{fig6}
(a) $\rho_a(T)$ data for $y$=6.45, 6.70, and 7.00 
in zero-field (thin lines) and in 16 T (thick lines).  
(b, c) Plots of $R_{H}(T)e/V_0$ and $S_a(T)$, respectively, 
for 60-K YBCO and LSCO at $x$=0.12.}
\end{figure}

It is worthwhile to note that the present data are useful 
for elucidating the origin of the 60-K plateau. 
Remember that the 60-K plateau is just a \textit{plateau} and is not a 
\textit{dip} in $T_c$, and there has been no clear evidence that the 
superconductivity is \textit{weakened} in the plateau; it should 
therefore be recognized that our result for $H_{c2}$ gives evidence 
that the superconductivity is actually weakened in the plateau. 
Since the weakening of the superconductivity is the fundamental feature 
of the 1/8 anomaly in the La-based cuprates, the present result clearly 
speaks for the 1/8-anomaly origin of the 60-K plateau.  
Furthermore, comparisons of the Hall coefficient $R_H$ and $a$-axis 
thermopower $S_a$ of 60-K YBCO to those of LSCO at $x$=0.12 
strongly suggest that the hole doping is in fact 1/8 for 
$y \simeq 6.65$; Figs. 6(b) and 6(c) show such comparisons in 
$R_{H}(T)e/V_0$ ($V_0$ is the volume per Cu in the plane) 
and $S_a(T)$, which demonstrate that LSCO at $x$=0.12 and YBCO at 
$y \simeq 6.65$ show quite similar values near room temperature, 
and these parameters have been proposed to be good indicators of the 
hole doping in cuprates \cite{Hanaki,Ong}.
Therefore, it appears that in the 60-K plateau the hole doping is 
actually $\sim$1/8 and the superconductivity is weakened, 
which together mean that the 60-K plateau is a manifestation of 
the 1/8 anomaly.  
It is useful to note that a reasonably good case for the connection 
between the plateau and the 1/8 anomaly was made previously in a 
complementary way, using Ca doping \cite{Ca-doping}.

It should however be noted that static charge stripes (a sort of 
charge density wave that localizes the carriers) do not appear to be 
the fundamental ``cause" of the weakening of the superconductivity 
in YBCO, since there is no evidence for static stripes in YBCO.
It is of course possible that the impact of the stripes at the 1/8 doping 
varies depending on the level of dynamics of the stripes in various 
cuprate systems. 
A more interesting possibility is that a proximity to a 
quantum critical point (QCP) \cite{QCP} is the more fundamental cause of 
the 1/8 anomaly: 
Recently, Aeppli \textit{et al.} reported \cite{Aeppli} that the 
magnetic response of the Nd-free LSCO (where there is also no evidence 
for static charge stripes) bears a signature of the quantum criticality.
In our data, fluctuations seem to be \textit{enhanced} near $y$=6.7 
while the superconductivity is \textit{suppressed}, which might also 
be an indication of a proximity to a QCP; if so, 
this particular QCP suppresses superconductivity through enhanced 
fluctuations, rather than creates superconductivity. 

Lastly, we note the implication of the overall evolution of $\xi_{ab}$.  
In cuprates, the pseudogap causes the energy gap $\Delta$ to \textit{grow} 
upon underdoping \cite{Timusk}, and thus there is no proportionality 
between $\Delta$ and $T_c$.  It is therefore not obvious whether 
$\xi_{ab}$ should follow the behavior of $\Delta$ or that of $T_c$. 
This is a rather fundamental problem in the pseudogap physics \cite{Levin}, 
but experimental data have been lacking.  The complicated $y$ dependence of 
$\xi_{ab}$ indicates that neither $\Delta$ nor $T_c$ are solely decisive 
and that an elaborate theory is necessary to describe $\xi_{ab}$.

We thank A. N. Lavrov, K. Levin, and N. P. Ong for helpful discussions, 
S. Komiya for providing LSCO crystals, and Y. Abe for technical 
assistance.

\end{document}